\pdfoutput=1 
\documentclass[a4paper]{article}

\usepackage[textwidth=16cm,textheight=24cm]{geometry}
\usepackage{amsmath}
\usepackage{amssymb}
\usepackage{amsfonts}
\usepackage{amsthm}
\usepackage{graphicx}
\usepackage{endnotes}
\usepackage{setspace}
\usepackage{verbatim}
\usepackage{times}
\usepackage{helvet}
\usepackage{courier}
\usepackage{bm}
\usepackage{url}
\usepackage{dcolumn}
\usepackage{cite} 
\usepackage[sort]{natbib}

\usepackage{multirow}
\usepackage{commath}

\usepackage[caption=true]{subfig} 
\usepackage[labelfont=it,textfont={sf}]{caption}
\usepackage{commath}

\title{
On the Origin of Risk Sensitivity: 
the Energy Budget Rule Revisited}

\author{
  Ik Soo Lim\\
  Bangor University, UK\\
  \texttt{i.s.lim@bangor.ac.uk}
  \and
  Peter Wittek\\
  University of Bor\aa s, Sweden\\
  \and
  John Parkinson\\
  Bangor University, UK
}

\date{}

\begin{document}
\maketitle

\begin{abstract}
  
The risk-sensitive foraging theory   
formulated in terms of the (daily) energy budget rule 
has been  influential in behavioural ecology
as well as other disciplines.
Predicting
 risk aversion on positive budgets
 and
 risk proneness on negative budgets,
however,
the budget rule has recently been challenged 
both empirically and theoretically.
In this paper,
we critically review these challenges
as well as the original derivation of the budget rule
and propose a `gradual' budget rule,
which is normatively derived from a gradual nature of risk sensitivity
and encompasses the conventional budget rule as a special case.
The gradual budget rule shows that the conventional budget rule holds 
when the expected reserve is close enough to a threshold for overnight survival,
selection pressure being significant.
The gradual view also reveals that the conventional budget rule does not need to hold
when  the expected reserve is not close enough to the threshold, 
selection pressure being insignificant.
The proposed gradual budget rule better fits the empirical findings
including those that used to challenge the conventional budget rule.

\vspace{0.5cm}


\end{abstract}

\section*{Introduction}

\subsection*{The Energy Budget Rule}
Life is full of risky decisions, from the humdrum to matters of life or death.
Understanding the drivers and the underlying processes of risky decisions
is important
in many disciplines including behavioural ecology.
\citet{caraco1980empirical} revolutionized foraging research by introducing risk sensitivity. The risk-sensitive foraging theory formulated in terms of the energy budget rule 
\citep{caraco1980empirical, stephens1981logic}
has been influential  
not only in behavioural ecology
\citep{kacelnik2013triumphs}, 
but also other disciplines
such as psychology, 
economics,
and anthropology
\citep{mishra2014decision, kalenscher2011why, mcdermott2008evolutionary, winterhalder1999risk}.
The  budget rule predicts that,
given  
a constant  feeding option and a variable option 
of equal mean,
an animal should choose the constant option if the expected energy budget is higher 
than the threshold required for overnight survival 
(i.e.\,the daily energy requirement)
and  the variable one if the expected budget is lower than the threshold;
i.e.\,risk aversion on positive energy budgets and risk proneness on negative budgets
\citep{stephens1981logic}.
The budget rule is logically appealing: it has an analytic derivation,
assuming that animals will behave to maximise the probability of avoiding  starvation~\citep{stephens1981logic}.
The budget rule has also been empirically supported~\citep{caraco1980empirical, caraco1990risk}.

\subsection*{Challenges against the Budget Rule}
In spite of the early successes and its  significant influence,
the budget rule has recently been challenged, both empirically and theoretically.
\citet{kacelnik2013triumphs} and~\citet{kacelnik1996risky} 
review the published empirical studies
related to the budget rule
and conclude that experimental support is weak.
One caveat to this conclusion is that
experiments
yielding negative results against the budget rule
tend to involve larger species
than those yielding positive results \citep{kacelnik1996risky}.
The $z$-score model which the budget rule is formally derived from assumes a small bird foraging
in order to meet daily energetic requirement 
in winter
\citep{stephens1981logic},
which is less critical for large species
\citep{caraco1990risk}.
Thus, any application of the budget rule to large species is less suitable  
since the short-term energetic requirement does not impose a significant threat to survival
unlike small species
\citep{caraco1990risk}.

More realistic models of risk sensitivity have been developed
\citep{mcnamara1996risk}, 
though one of the main drawbacks with these models is 
that their predictions are less testable due to the increased complexity
\citep{kacelnik2013triumphs}.
\citet{kacelnik2013triumphs}
assert
that
risk sensitivity does not have adaptive value by itself
and it is a side-effect 
caused by psychophysical features of animal information processing systems. 
Scalar Utility Theory is proposed,
 based on Weber's law
that relates the perceived difference between stimuli to the mean value of the stimuli~\citep{kacelnik2013triumphs, kacelnik1996risky}.
However, Scalar Utility Theory predicts preference of risk aversion alone 
(when variability is in the amount of food) 
whereas there are empirical findings that report preference of risk proneness,
which cannot be ignored.
More importantly, 
the mechanistic (proximate) explanation based on the psychophysical features
does not need to deny the functional (ultimate) explanations 
since they are complementary rather than competing alternatives.
Based on a meta-analysis of
the same studies listed in~\citet{kacelnik1996risky}
as well as more recent ones,
for instance,~\citet{shafir2000risk}
draws a conclusion different from that of~\citet{kacelnik2013triumphs} and~\citet{kacelnik1996risky}.
While applying Weber's law as well,~\citet{shafir2000risk} presents regression-based results
that accommodate both risk aversion and risk proneness
to be compatible with the budget rule,
which 
integrates the  functional and mechanistic explanations;
the direction of preference is determined by functional considerations (the budget rule)
and the strength
of preference by perceptual considerations (Weber's law).
Rather than exclusively deducing either risk sensitivity or risk indifference
per study,
\citet{shafir2000risk} examines the `degree' of risk sensitivity
from a perceptual perspective 
and
shows that 
there is a gradient of responses to risk.

In this paper,
we present a gradual view of risk sensitivity from a normative perspective
and provide a gradual version of the budget rule,
which not only explains the empirically observed gradual nature of risk sensitivity,
but also can explain observations counted against the budget rule.

\section*{Critical Review of the Derivation of the Budget Rule}

We review the 
formal derivation of the budget rule
from the $z$-score model
\citep{stephens1981logic, stephens1987foraging}
and point out what it overlooks.
The $z$-score model assumes a small bird foraging in winter 
to survive a night during which it cannot forage.
We adopt notations similar to those of~\citet{stephens1987foraging}.
The bird's energy reserve $x$
at dusk
 is assumed to be normally distributed
with  mean $\mu$ and variance $\sigma^2$.
It is also assumed
that the bird has behavioural control
over $\mu$ and $\sigma$,
having to 
meet a fixed daily energetic requirement  $R$ 
to survive the night.
Fitness is proportional to short-term survivorship
which is represented 
 by a step function in the $z$-score model,
yielding the expected fitness to be proportional to the survival probability $f$.
The model seeks to maximise the survival probability 
\begin{eqnarray*}
f &=& \Pr(\text{surviving the night}) \\
&=& \Pr\left(x > R\right) \\
	       &=&  1 - \Phi \left( z \right) \label{cum_dis}
\end{eqnarray*}
where 
$z=\frac{R-\mu}{\sigma}$ is a $z$-score and 
$\Phi (\cdot)$
is the cumulative distribution function
of the standard 
normal distribution
$\mathcal{N}_{0,1}(\cdot)$
with zero mean and unit variance.
    To maximise $f$
for a given $\mu$,
one needs to check its first derivative
\begin{eqnarray*}
  \dpd{f}{\sigma}
   &=& - \dpd{\Phi}{\sigma} \\
	       &=&   -   \dod{\Phi}{z} \dpd{z}{\sigma} \label{eq_rate_change}
\end{eqnarray*}
where $ \pd{z}{\sigma}  = \frac{\mu - R}{\sigma^2}$.
Note that the sign of $\pd{f}{\sigma}$
is entirely determined by 
that of
$\pd{z}{\sigma} $ 
in the sense that
$\od{\Phi}{z} >0$  always holds;
if $z$ decreases, then
survival probability $f$  increases and vice versa.
The analytical
derivation of the  budget rule
then entirely focuses on the sign of $\pd{z}{\sigma} $
\citep{stephens1987foraging}.
Given a constant mean and different variances,
it yields
the effect of $\sigma$ on $z$ as follows;
\begin{eqnarray*}
\dpd{z}{\sigma}    & >  & 0 \text{\hspace{0.5cm} if } \mu > R, \label{eq_sign_z} \\ 
\dpd{z}{\sigma}     & <  & 0 \text{\hspace{0.5cm} if } \mu < R.
\end{eqnarray*}
On positive budgets $\mu > R$,
the standard deviation $\sigma$ should be reduced
to decrease $z$.
On negative budgets  $\mu < R $,
$\sigma$ should be increased
to decrease $z$.
This yields the extreme variance rule,
predicting that the smallest variance possible is the optimal behaviour to be chosen on positive budgets
and the largest variance possible
on negative budgets
\citep{stephens1987foraging}.
The budget rule is a special case of the extreme variance rule
with the smallest variance  being zero,
predicting the optimal risk sensitivity of 
risk aversion on positive budgets
and risk proneness on negative budgets.

A caveat of 
this 
and its standard interpretation
is
an all-or-nothing view on
animal decision-making,
focusing mainly on
optimal behaviour.
According to this all-or-nothing view,
if an animal's response to variability does not comply with the optimal risk sensitivity
(in a statistically significant sense),
it is counted against the budget rule
\citep{kacelnik2013triumphs,kacelnik1996risky}.
However,
comparison of observed and predicted behaviours is not sufficient
for testing a behavioural theory
and one should not reject the theory without comparing the fitness of them
\citep{mangel1991adaptive}.
Not all of the observations different from the prediction
should be treated the same.
If the difference in fitness
between 
observed and predicted behaviours
is insignificant,
for instance,
one should not use the observation to reject the theory.
The degree of risk sensitivity under different experimental conditions is thus more informative 
than merely testing whether the response to risk complies with the prediction of the budget rule;
according to the meta-analysis of
the same empirical studies listed in~\citet{kacelnik1996risky},
animals indeed show 
a gradient of responses to risk,
the degree of risk sensitivity forming a continuum
\citep{shafir2000risk}.
While the conclusion of \citet{shafir2000risk} is based on different groups of subjects under different energy budgets,
the gradual nature of risk-sensitive responses within the same individuals
is also observed \citep{hurly2003twin,bacon2010both}. 
Thus, a more gradual view of the budget rule 
needs to be sought.
Another caveat  of the derivation
is
the $z$-score centric view.
Since
the $z$-score has an unlimited range of $-\infty < z < \infty$
whereas the survival probability has a limited range of $0 \le f(z) \le 1$
with a one-to-one correspondence between $z$ and $f(z)$,
a large change in $z$ can yield little change in $f(z)$.
If foraging options significantly differ in $z$,
a $z$-score centric view
can misleadingly imply a strong risk sensitivity
whereas there is actually a weak risk sensitivity or a risk indifference 
due to little difference in corresponding survival probability $f(z)$.

\section*{Gradual Nature of Risk Sensitivity}
There already exists a theoretical consideration on a gradual nature of risk sensitivity.
\citet{caraco1985foraging} formulated decreasing risk aversion
of the $z$-score model 
based on the marginal rate of substitution
\citep{keeney1976decisions},
\begin{equation*}
\frac{\dpd{z}{(\sigma^2)}}{\dpd{z}{\mu}}= \frac{R-\mu}{2 \sigma^2} = \frac{z}{2\sigma}.
\label{eq_local_risk_sensitivity}
\end{equation*}
A caveat of this analysis is that the gradual nature of risk sensitivity is considered 
only 
along an indifference curve of a constant $z$-score
equivalent to a constant survival probability;
the analysis varies mean $\mu$ and variance $\sigma^2$ such that a constant $z$-score is maintained.
This is less compatible with
the standard design of risk sensitivity experiments
that offer a choice 
between a variable ($\sigma \ne 0$) and a constant ($\sigma=0$) feeding option with the same mean,
yielding the options to differ in $z$-score.

The Arrow-Pratt measure 
popular in economics~\citep{pratt1964risk}
\begin{equation*}
\rho=-\frac{u''(x)}{u'(x)} 
\end{equation*}
is often used 
as a risk aversion measure
where $u(x)$ is a 
utility (or fitness) function
\citep{real1986risk}.
Decreasing risk aversion is indicated by
$\rho' =\od{\rho}{x}< 0$.
A caveat is that it is a local measure.
It only informs of the response to risk at a single point along the fitness function
whereas risk sensitivity deals with problems
that require more than information at a single point to predict preference,
probability distributions spreading likelihood over many possible amounts
\citep{stephens1987foraging}.
Moreover,
the Arrow-Pratt measure for the $z$-score model is not well defined
since
$u'(x)=u''(x)=0$ due to the fitness function $u(x)$ being a step function 
(i.e.\,piece-wise constant).

These caveats of the existing measures call for an alternative measure of gradual nature of risk sensitivity,
which should satisfy the following requirements: 
\begin{enumerate}
\item[(1)] It should reflect risk sensitivity 
in a choice between a variable and a constant feeding 
option
with the same mean.
\item[(2)] It should take into consideration the whole distribution of reserve rather than a single value of it.
\end{enumerate}

\subsection*{Measure of Risk Sensitivity}

We propose a  measure of risk sensitivity from a normative perspective
 \begin{equation*}
 S =  \frac{U (\mu,\sigma) - U(\mu,0)}{{\triangle U}_{\max}} 
\end{equation*}
  where $U(\mu,\sigma)$ is the expected fitness of a feeding option with mean $\mu$ and variance $\sigma^2$,
  and 
  ${\triangle U}_{\max}$ is the maximum possible difference in $U$.
 It satisfies the aforementioned two requirements.
It takes  into consideration the whole distribution of reserve $x$ at dusk
in the sense that 
rather than fitness $u(x)$ at a single value $x$,
it uses expected fitness $U(\mu,\sigma)= \int \mathcal{N}_{\mu,\sigma}(x) u(x) dx$
obtained by averaging or integrating $u(x)$
with the normal distribution  $\mathcal{N}_{\mu,\sigma}(x)$ of reserve $x$.
It also reflects a choice between a variable and a constant feeding option with the same mean
since it involves the expected fitness $U(\mu,\sigma)$ of the variable option
and $U(\mu,0)$ of the constant option. 
The reason why animals should show risk sensitivity in a normative sense is 
a consequence in expected fitness,
which is well reflected in the proposed measure $S$ of risk sensitivity;
the greater difference in expected fitness 
$U(\mu,\sigma) - U(\mu,0)$
between the options, 
the greater sensitivity of risk should animals show.
For instance,
the greater is the expected fitness $U(\mu,\sigma)$
of the variable option
than $U(\mu,0)$ of the constant option
(i.e.\,the greater is $S$),
the greater is selection pressure
for preferring the variable option (i.e.\,the greater is risk proneness).
 Note that the proposed measure $S$ is invariant 
 under an affine transformation of the expected fitness $U$.
 In other words,
 even if $U$ is transformed into $Y = a U + b$  via scaling by $a$ and offsetting by $b$, the risk sensitivity measured by $S$ remains the same,
$S =  \frac{U (\mu,\sigma) - U(\mu,0)}{{\triangle U}_{\max}} =\frac{Y (\mu,\sigma) - Y(\mu,0)}{{\triangle Y}_{\max}}$.
This affine-invariance of $S$ is 
 due to the division by ${\triangle U}_{\max}$
 and the subtraction $U(\mu,\sigma) - U(\mu,0)$. 
One of the reasons
 why the Arrow-Pratt measure  
involves the division by $u'(x)$
 is to maintain the affine-invariance.
The affine-invariance 
of $S$  
is important;
 for instance, the animal's internal representation may not use expected fitness $U$ directly, 
 but its affine transformation $Y$.

Since the expected fitness in the $z$-score model is proportional to the survival probability $f$
and ${\triangle f}_{\max}=1$,
we have 
 $ S =  \frac{f (\mu,\sigma) - f(\mu,0)}{{\triangle f}_{\max}} = f(\mu,\sigma) - f(\mu,0)$
due to the affine-invariance;
 note that 
 the maximum possible difference in $f$ is ${\triangle f}_{\max}=1$
 since probability $f$ ranges from 0 to 1. 
 The difference in survival probability
$\triangle f(\mu, \sigma) = f(\mu, \sigma) - f(\mu, 0)$
between a variable and a constant foraging 
option 
with the same mean $\mu$
thus serves as a measure of risk sensitivity
that an animal should show
due to a consequence in expected fitness.
The sign of 
 $\triangle f$ can discriminate between risk aversion and risk proneness.
 $\triangle f < 0$ implies  selection  for risk aversion  since the survival probability of the constant option $f(\mu, 0)$ is higher than that of the variable option $f(\mu, \sigma)$ whereas
 $\triangle f > 0$ implies selection  for risk proneness. 
 The magnitude of $ \triangle f $ indicates strength of selection for risk sensitivity;
 the larger $\left| \triangle f \right|$, the stronger selection pressure for risk sensitivity.
For instance,
the larger is $\left| \triangle f \right|$
while $\triangle f < 0$,
the stronger is the selection pressure for risk aversion;
thus, the more frequent choice of the constant option
over the variable one.

\subsection*{Formal Derivation of Gradual Risk Sensitivity}
With the measure of risk sensitivity $\triangle f(\mu, \sigma)$,
we now derive the gradual nature of risk sensitivity from the $z$-score model.
We are particularly interested in the degree of risk sensitivity
with respect to the energy budget.
In other words,
for a given variable option $\sigma$ (and the other option being a constant one),
how does the degree of risk sensitivity vary as mean reserve $\mu$ varies?
For this,
we need to compute $\pd{|\triangle f |}{\mu}$
that is the rate of change in (the magnitude of) risk sensitivity $|\triangle f |$
as $\mu$ increases
for a given $\sigma$.
Especially,
the sign of $\pd{|\triangle f |}{\mu}$ is useful;
$\pd{|\triangle f |}{\mu} < 0$ indicates decreasing risk sensitivity
whereas $\pd{|\triangle f |}{\mu} > 0$, increasing risk sensitivity. 
 
\begin{figure*}[t]
\centering
\includegraphics[width=3in]{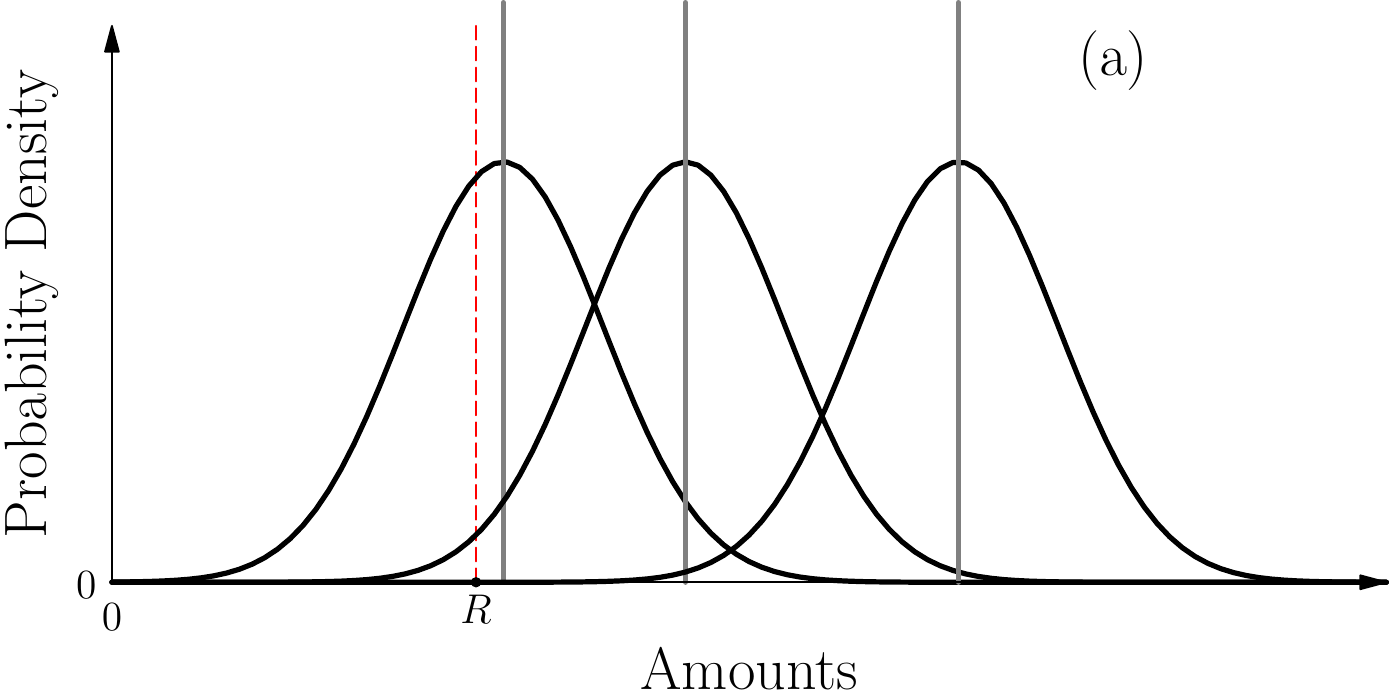}
\hspace{0.5cm}
\includegraphics[width=3in]{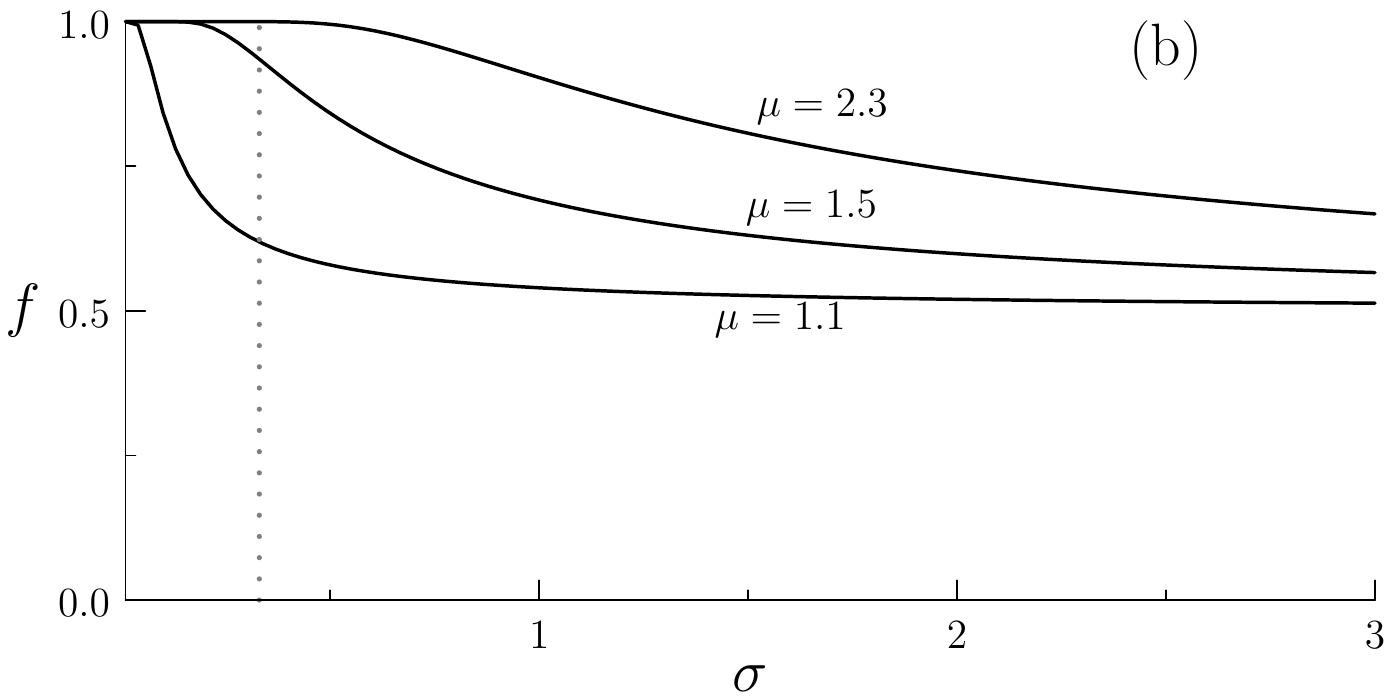}
\caption{ 
Gradual nature of risk sensitivity.
	(a)
Three pairs of distributions for a variable ($\sigma \ne 0$) and a constant ($\sigma=0$)  foraging option are shown, 
each pair having an equal mean.
The overnight survival threshold $R$ is indicated by a dashed line.
For a given variance $\sigma^2$, 
the difference in survival probability 
between a variable and a constant option 
changes,
depending on the mean $\mu$ of each reserve distribution.
Note that the survival probability is equal to the area under the distribution graph beyond $R$.
As the mean of a distribution increases on positive budgets,
the difference in survival probability between a pair of the options decreases 
from $\frac{1}{2}$ to $0$.
In other words,
the strength of selection 
for risk aversion gradually decreases and can diminish
as the mean increases;
i.e.\,decreasing and diminishing risk aversion. 
For instance,
the distribution 
for the variable option
that is the furthest away from the threshold
has the survival probability close to 1
while the probability for the constant option is 1
and so the difference is close to 0;
i.e.\,weak selection for risk aversion.
The further away from the threshold,
the weaker
selection for risk sensitivity.
		(b)  
The survival probability $f$ 
as a function of standard deviation $\sigma$ of the reserve distribution
with mean $\mu=1.1, 1.5, \text{ and } 2.3$ and threshold $R=1$ as those in (a).
As $\mu$ increases for a given $\sigma$,
$f$ increases and approaches to 1,
the optimal probability obtained at $\sigma=0$; 
for example,
 $\sigma=0.33$ used for the three distributions in  (a) is indicated by a dotted line.
The higher $\mu$,
the wider range of $\sigma$ yielding $f$ close to 1.
}
\label{fig_risk_sensitivity}
\end{figure*}

On positive budgets $\mu>R$,
we have
$\triangle f(\mu, \sigma) 
= f(\mu, \sigma) -  f(\mu, 0)  = f(\mu, \sigma) - 1 < 0$
~(i.e. risk aversion)
since the constant option guarantees the survival (i.e.\,$f(\mu,0)=1$)
whereas the variable one cannot (i.e.\,$f(\mu, \sigma) <1$).
Since 
$\left|\triangle f(\mu, \sigma)\right| 
= -\left(f(\mu, \sigma) - 1 \right)$,
we have
$\pd{|\triangle f |}{\mu} = -\pd{f}{\mu}$.
For a given $\sigma$,
we 
have
\begin{eqnarray*}
\dpd{f}{\mu}
	       &=&   - \dod{\Phi}{z} \dpd{z}{\mu} \\
	       &=&   \frac{1}{\sqrt{2\pi}}\exp\left(-\frac{\left(R -\mu \right)^2}{2 \sigma^2}\right)\frac{1}{\sigma}
\end{eqnarray*}
Note that $\pd{f}{\mu} > 0$ always holds for  $\sigma > 0$
and thus $\pd{|\triangle f |}{\mu} < 0$.
In other words,
the strength of selection for risk aversion
or the degree of risk aversion
$ |\triangle f | $ decreases as $\mu$ increases on positive budgets
for a given $\sigma$.
Moreover,
we have
$f(\mu, \sigma)  \rightarrow 1$ as $\mu$ increases for a given $\sigma$
since 
$f = 1 - \Phi \left( z \right) $ and $\Phi \left( z \right)  \rightarrow 0$
as  $z=\frac{R-\mu}{\sigma} \rightarrow -\infty$.
In other words,
the survival is almost guaranteed even if the variable option is chosen
as long as the reserve mean is high enough.
Hence,
we have
$ |\triangle f | \rightarrow 0$ as $\mu$ increases for a given $\sigma$;
i.e.\,there is little difference in the survival probability between the feeding options
if the reserve mean is high enough.
The conclusion is that on positive budgets,
the strength of selection 
for risk aversion monotonically decreases and can diminish
as the reserve mean increases for a given variance.
This gradual and diminishing nature of risk aversion  can be intuitively understood as well;
see Fig.\,\ref{fig_risk_sensitivity}.

\begin{figure*}[t]
\centering
\includegraphics[height=5cm]{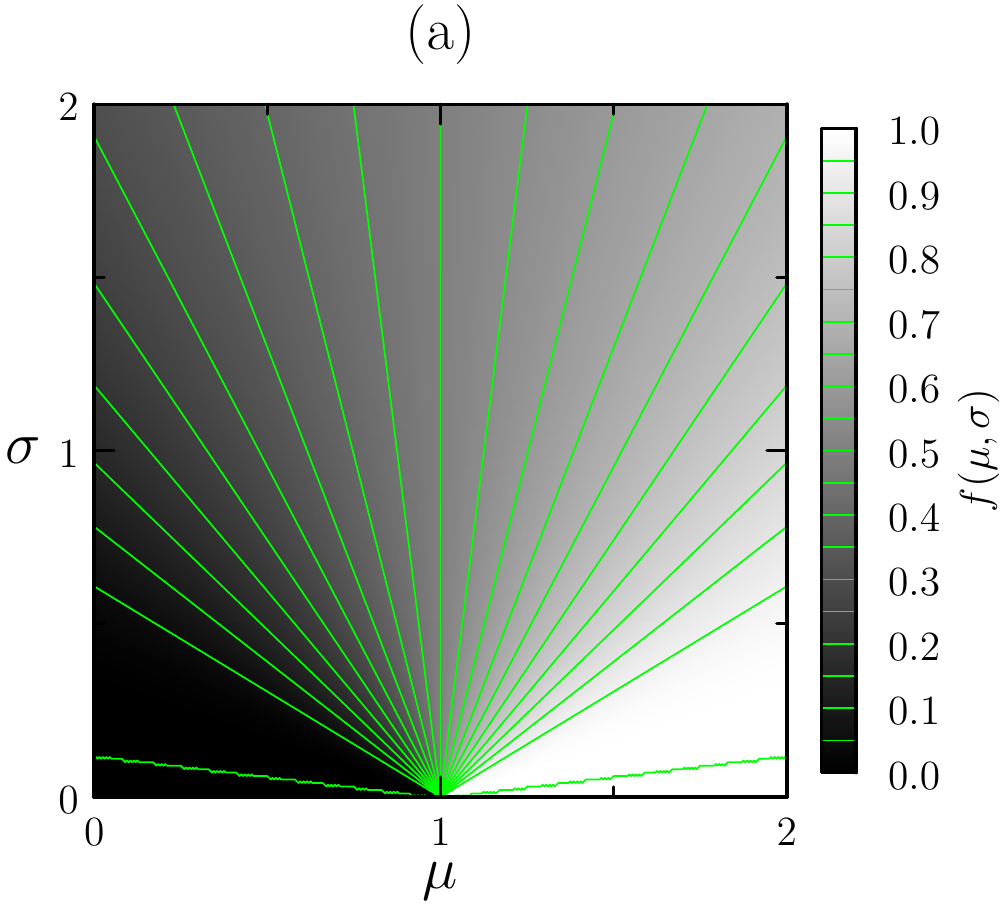}
\hspace{1cm}
\includegraphics[height=5cm]{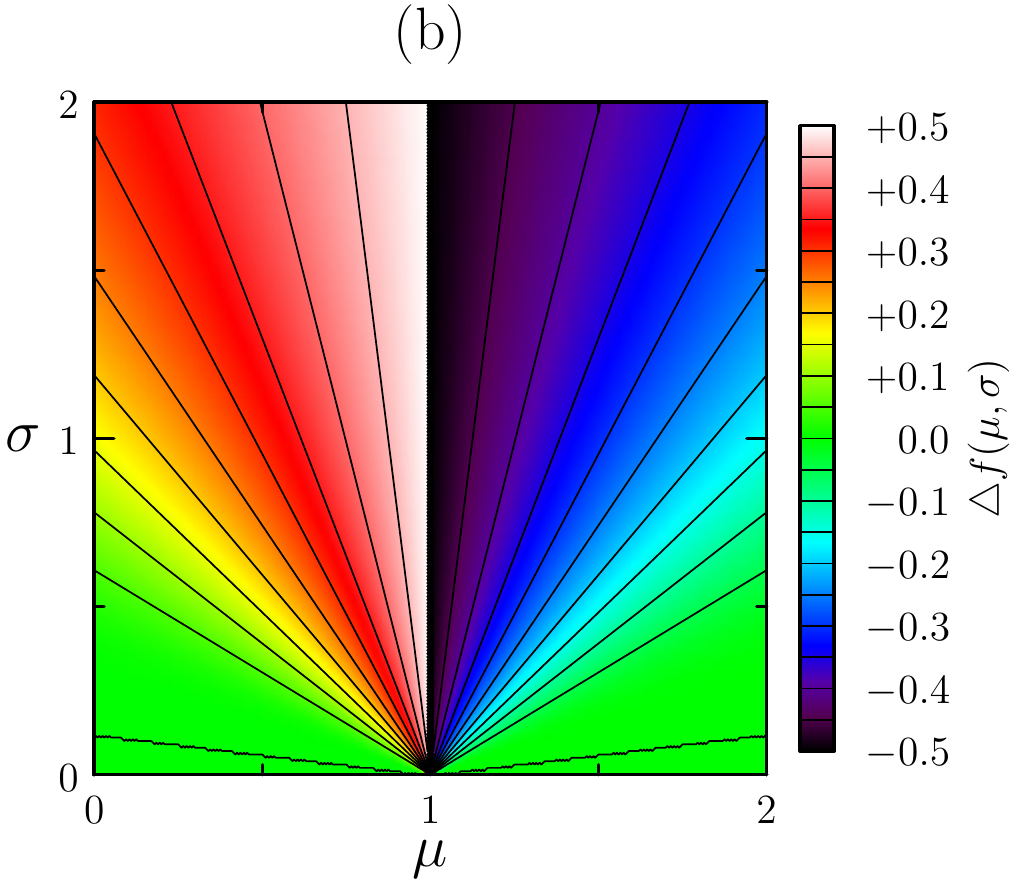}
\caption{ 
(a)
The values of the survival probability  $f(\mu, \sigma)$ with the threshold $R=1$
are plotted as gray-level intensities
on the $\mu - \sigma$ plane.
Contours  
corresponding to $f=0.05, 0.1, \cdots,0.9, 0.95$
are also plotted. 
(b)
The probability residuals
$\triangle f(\mu, \sigma) = f(\mu, \sigma) - f(\mu, 0)$ 
are plotted in colour.
Showing the difference in survival probabilities between variable and constant foraging options,
the residual indicates  
levels of selection pressure for risk sensitivity.
Contours at $\triangle f= -0.5, -0.45,  \cdots,0.45, 0.5$.
The larger $\left| \triangle f \right|$ with $ \triangle f > 0$, the stronger selection pressure for risk proneness (red-white);
the larger $\left| \triangle f \right|$ with $ \triangle f < 0$, the stronger selection pressure for risk aversion (blue-black);
$\triangle f \approx 0$, weak selection pressure for risk sensitivity (green).
}
\label{fig_iso_prob}
\end{figure*}

On negative budgets,
by a similar argument,
the strength of selection for risk proneness monotonically decreases and can diminish
as the mean decreases for a given variance;
see Fig.\,\ref{fig_iso_prob}.
In other words,
the degree of  optimal risk sensitivity predicted by the conventional budget rule
gradually changes. 
Being purely based on an adaptive rationale,
this `gradual' view of the $z$-score model
(or the gradual budget rule hereafter) predicts that 
the conventional budget rule holds 
if the difference in 
expected fitness
(or survival probability) between the feeding options
is significant 
(e.g.\,when the reserve mean is close to starvation threshold for a given deviation)
and may not do so otherwise,
risk sensitivity gradually weakening.
The gradual budget rule not only encompasses the  conventional budget rule as a special case of it,
but also may explain empirical results that are often counted against
the conventional budget rule
 and  the adaptive rationale of risk sensitivity behind it.

\subsection*{Weak Selection Pressure vs.\,Weak Risk Sensitivity}

The empirical findings
that report 
risk indifference
or
weak risk sensitivity
used to be counted against
the conventional budget rule
\citep{kacelnik2013triumphs, kacelnik1996risky}. 
According to the gradual budget rule,
however,
if the difference in expected fitness
 between 
the feeding options
is insignificant,
risk indifference is feasible
in the sense that
there is weak selection for the optimal risk sensitivity
predicted by the conventional budget rule
and, thus,
weak selection against risk indifference.
Note that weak selection for the optimal risk sensitivity 
does not imply strong selection for risk indifference,
the latter of which would predict only risk indifference.
Instead,
 it implies weak selection against  suboptimal risk sensitivity as well as risk indifference
in the sense that
there is little difference in expected fitness between the optimal and the suboptimal risk sensitivity.
In other words,
risk aversion, risk proneness and risk indifference are all feasible
under the condition of weak selection,
which may explain empirically observed risk indifference and suboptimal risk sensitivity
such as risk proneness on positive budgets and risk aversion on negative budgets.
Thus,
one should not simply conclude that 
any observation apparently contradicting the optimal risk sensitivity
challenges the conventional budget rule
 and  the adaptive rationale of risk sensitivity behind it.
Instead, one should take into consideration the expected fitness of the observation
and  that of the optimal risk sensitivity.

\subsubsection*{Risk proneness on positive budgets and risk aversion on negative budgets}

Other than the argument of weak selection,
can we provide additional explanations
why risk sensitivity (especially, suboptimal risk sensitivity such as risk proneness on positive budgets
and risk aversion on negative budgets)
 rather than risk indifference
can be empirically observed in a statistically significant sense?
Any model is an approximation or simplification of the reality, leaving out many features.
The $z$-score model that the budget rule is formally derived from
incorporates the survival,
but leaves out other functional and 
mechanistic 
features 
such as reproduction and perception.
Under the condition of weak selection for the optimal risk sensitivity
(e.g.\,the reserve mean is distant enough from the survival threshold for a given variance),
any of risk proneness, risk aversion and risk indifference is feasible
from the gradual view of the $z$-score model.
However,
only one of them may be feasible from the perspective of a feature that the $z$-score model leaves out, say, reproduction;
e.g.\,only risk proneness may yield expected fitness substantially higher than that of any other alternatives in terms of reproduction.
We elaborate this aspect.

\begin{figure*}[t]
\centering
\includegraphics[height=4.8cm]{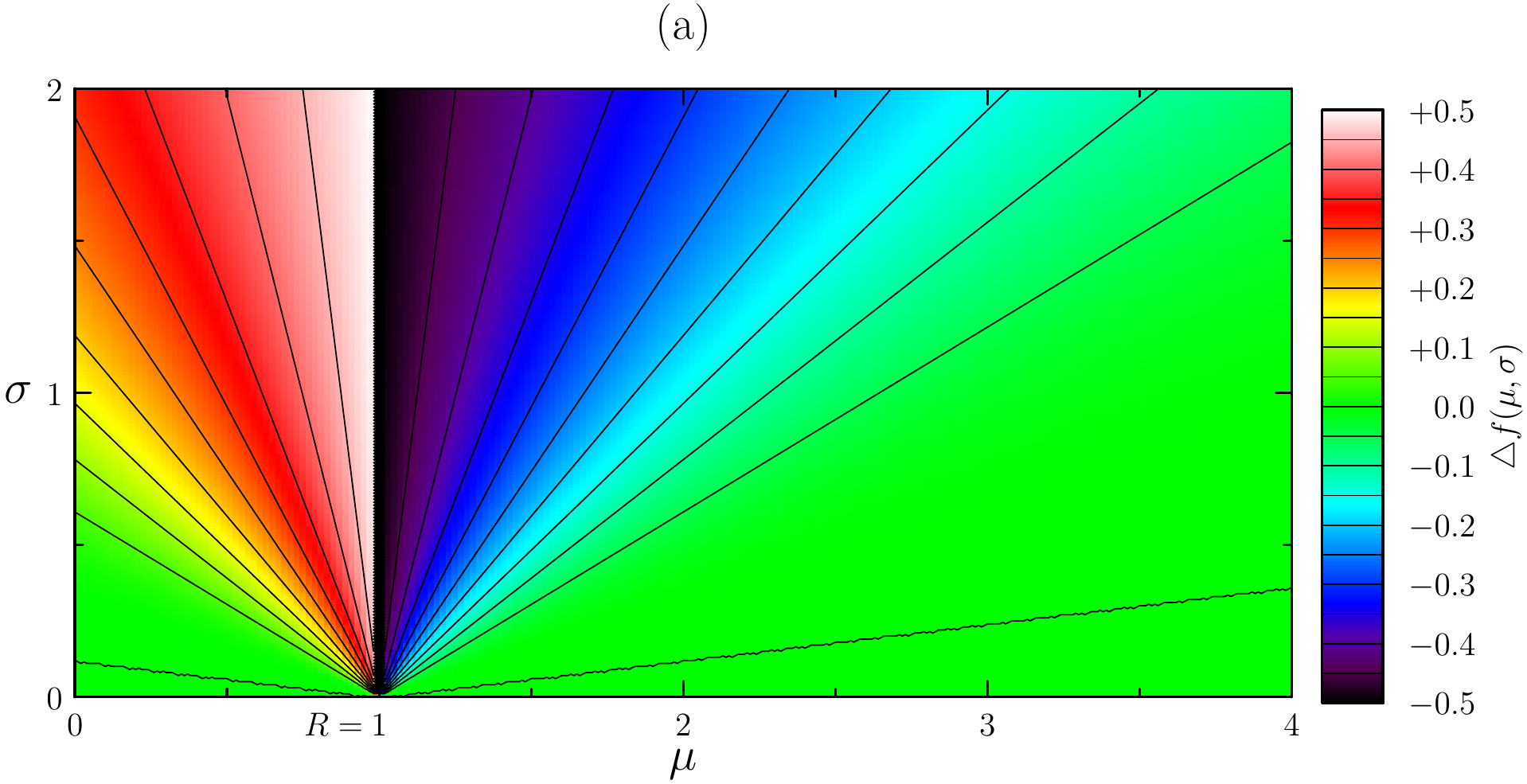}    
\hspace{2cm}
\includegraphics[height=4.8cm]{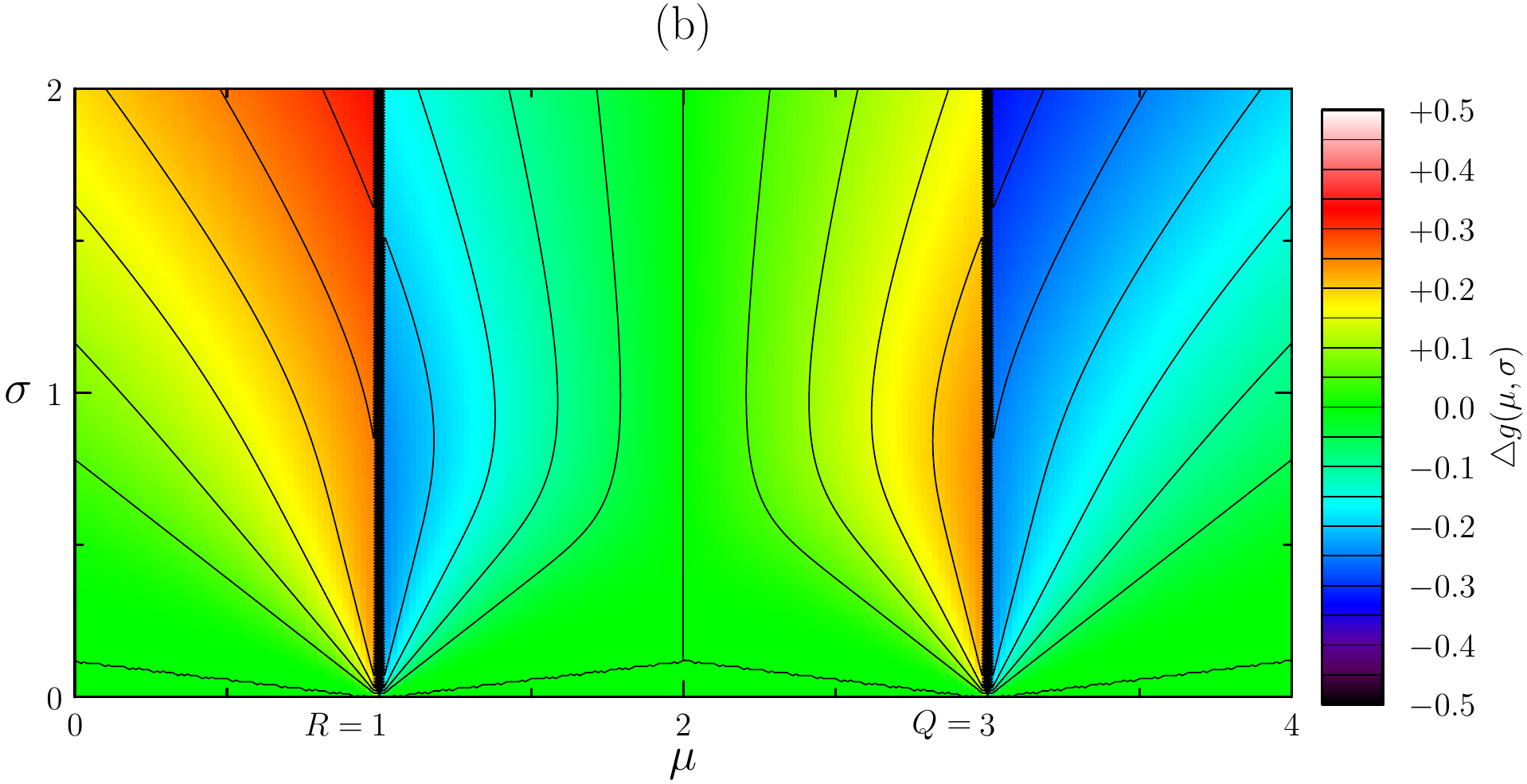}
\caption{ 
Comparison of gradual risk sensitivity
between the $z$-score model  and the twin threshold model.
(a) 
The gradual risk sensitivity $\triangle f$ from the $z$-score model
with  $R=1$.
For a given standard deviation $\sigma$,
the degree of risk sensitivity (risk aversion) $| \triangle f |$ on positive budgets $\mu > R$ gradually decreases and diminshes
as mean $\mu$ increases,
which can yield any of risk proneness and risk indifference as well as risk aversion
due to weak selection pressure,
for instance, $\triangle f \approx 0.0$ near at  $(\mu, \sigma)=(2.9, 1.0)$.
(b)
The gradual risk sensitivity $\triangle g$ from the twin threshold model
with $R=1$,  $Q=3$, and $w=\frac{1}{2}$.
For a given $\sigma$,
the degree of risk sensitivity (risk aversion) $| \triangle g|$  on positive budgets gradually decreases and diminishes
as  $\mu$ increases up to $\mu=2$
and then the degree of risk sensitivity (risk proneness) $| \triangle g|$
gradually increases 
as $\mu$ increases up to $\mu=Q$,
the latter of which predicts only risk proneness,
refining the prediction based on $\triangle f$,
for instance, 
$\triangle g \approx 0.25$ near at  $(\mu, \sigma)=(2.9, 1.0)$.
}
\label{fig_twin} 
\end{figure*}

Being built upon the $z$-score model,
the twin threshold model
seeks to maximise the combined probability $g$ of surviving the night and succeeding in reproduction
\citep{hurly2003twin}
\begin{eqnarray*}
g (\mu,\sigma) &=& w \Pr(\text{surviving the night}) + (1-w) \Pr(\text{succeeding in reproduction}) \\
&=& w \Pr\left(x > R\right) + (1-w) \Pr\left(x > Q\right)  \\
	       &=&   w \left( 1 - \Phi \left( \frac{R-\mu}{\sigma} \right)\right) + (1-w) \left( 1 - \Phi \left( \frac{Q-\mu}{\sigma} \right)  \right) 
\end{eqnarray*}
where $Q$ is reproduction threshold, $R < Q$ and $ w \in [0,1]$.
Whereas the original twin threshold model uses  combination weights $w$ and 1,
we use the convex combination of the two probabilities with $w$ and $1-w$
to ensure that the combination $g (\mu,\sigma)$ remains a probability.
With starvation threshold $R$ and reproduction threshold $Q$,
their associated residual or difference in 
probability
$\triangle g (\mu,\sigma) = g (\mu,\sigma) - g (\mu,0)$
is qualitatively different from the residual $\triangle f (\mu,\sigma)$ associated with a single threshold $R$;
see Fig.\,\ref{fig_twin}.
As empirically observed from some species,
the twin threshold model predicts risk proneness on positive budgets
(i.e.\,$\triangle g>0$ being significant)
in a range of parameters $(\mu, \sigma)$
where reserve mean $\mu$ is 
close to but lower than reproduction threshold $Q$.
This makes sense since risk proneness offers a possibility of reproduction
whereas risk aversion does not.
In the same range of parameters,
the gradual budget rule
can yield weak selection pressure for the optimal risk sensitivity predicted by the conventional budget rule
(i.e.\,$\triangle f$ being insignificant)
and so
predicts
 that 
any of  risk proneness, risk aversion or risk indifference is feasible,
there being little difference in expected fitness between them.
Near at  $(\mu, \sigma)=(2.9, 1.0)$,
for instance,
we have
$\triangle g \approx 0.25$ whereas $\triangle f \approx 0.0$
in Fig.\,\ref{fig_twin}.
Rather than conflicted predictions,
the inclusion of reproduction feature
refines the predictions based on only survival feature
in the sense that the latter predicts any of risk proneness, risk aversion or risk indifference to be feasible
and the first predicts only one of them (i.e.\,risk proneness) to be feasible.
Note that they make similar predictions 
when both of the models are under conditions of strong selection,
i.e.\,near starvation threshold $R$.
As  reserve mean $\mu$ is lowered and becomes close to starvation threshold $R$ on positive budgets
for a given  variance $\sigma^2$,
for instance,
selection pressure for risk proneness to succeed in reproduction weakens  
whereas selection pressure for risk aversion to survive the night strengthens
and, thus,
the gradual view of the twin threshold model predicts risk aversion 
as 
the gradual view of the $z$-score model does.
Near at  $(\mu, \sigma)=(1.1, 1.0)$,
for instance,
we have $\triangle g \approx - 0.25$ and $\triangle f \approx -0.4$ in Fig.\,\ref{fig_twin},
both yielding strong selection pressure for risk aversion. 

Another feature that the $z$-score model leaves out is 
the possibility of 
starvation (to death) during foraging,
which can explain risk aversion on negative budgets.
A forager (e.g. a small bird) may have to ensure it stays above a lethal boundary of reserves during foraging as well,
which is lower than the overnight starvation threshold $R$.
There are models 
that take into consideration
both possibilities of starvation during foraging and night,
yielding a prediction different from that of the $z$-score model
\mbox{\citep{houston1985choice,stephens1987foraging,stephens1982stochasticity}}.
For instance,
the lazy-\textit{L} model predicts risk aversion on negative budgets 
if the requirement to avoid the lethal boundary
is a lot stronger than the requirement to survive the night
\mbox{\citep{stephens1987foraging,stephens1982stochasticity}}.
This is especially so when there is enough time left for foraging until dusk
and reserves are dangerously low.
As dusk approaches,
however,
the requirement to survive the night
strengthens
whereas the effect of the lethal boundary weakens,
which makes
the lazy-\textit{L} model become more equivalent to
 the $z$-score model~\citep{stephens1987foraging,stephens1982stochasticity};
increasing risk proneness on negative budgets is thus predicted
by both models
as reserve mean increases.

In summary,
additional features that the $z$-score model leaves out
can refine
predictions 
made by 
the gradual budget rule
under conditions of weak selection,
which may explain risk proneness on positive budgets and risk aversion on negative budgets,
contrary to the predictions of the conventional budget rule.

\section*{Discussion}

We derived the gradual budget rule
that normatively reflects the gradual nature of risk sensitivity
and encompasses the conventional budget rule as a special case.
The gradual budget rule better fits the empirical findings including those that used to challenge the conventional budget rule.
The proposed measure of gradual risk sensitivity $S =  \frac{U (\mu,\sigma) - U(\mu,0)}{{\triangle U}_{\max}}$
is only applicable to a choice between
a variable ($\sigma \ne 0$) and a constant feeding option
($\sigma=0$)
with the same mean $\mu$.
We can relax some of the conditions on $S$
and generalise it 
for a wider range of choices.
Instead of insisting on one of the feeding options to be a constant one,
we may consider that both options are variable
and we then have $S_{\sigma} =  \frac{U (\mu,\sigma_1) - U(\mu,\sigma_2)}{{\triangle U}_{\max}}$.
Note that $S$ is a special case of $S_{\sigma}$ with $\sigma_2=0$.
We may further relax the condition that both options should have the same mean $\mu$
and we then have
$S_{\mu,\sigma} =  \frac{U (\mu_1,\sigma_1) - U(\mu_2,\sigma_2)}{{\triangle U}_{\max}}$.
Note that 
$S_{\sigma}$ is a special case of $S_{\mu,\sigma}$ with $\mu_1=\mu_2=\mu$
while
 $S$
 is a special case with $\mu_1=\mu_2=\mu$
and $\sigma_2=0$.
In the remaining  part of this section,
we critically compare and discuss other relevant work.

\subsection*{Partial Preference and Behavioural Landscapes}
A metric relevant to the gradual risk sensitivity  is the canonical cost of a suboptimal behaviour
$c=U^*-U$  where $U^*$ is the expected fitness for the optimal behaviour
and $U$ for a suboptimal behaviour
\citep{mcnamara1986common}.
A caveat of this metric is that it is not affine-invariant.
For instance,
the expected fitness scaled by $a \ne 0$ fails the invariance,
i.e.\,$U^*-U \ne aU^* - aU$;
rather than expected fitness itself
(i.e.\,the expected number of offspring surviving to adulthood),
note that a scaled expected fitness is often used
\citep{nevai2007state}.
As a function of the canonical cost $c$,
the probability of  choosing a suboptimal behaviour can be derived,
assuming that the perception of expected fitness $U$  is subject to error
\citep{mcnamara1987partial}.
When  an animal makes a choice between two simultaneously available options,
one has
$ \pi(c) = \frac{\exp(-\beta c)}{1 + \exp(-\beta c)}$
for  the probability of a suboptimal option
and 
$1 - \pi(c)$
for the optimal option
where $\beta$ is a positive free parameter
\citep{houston1997natural}.
It yields $\pi(c) \approx \frac{1}{2}$
when $c \approx 0$,
predicting only risk indifference 
when there is little difference in expected fitness between the optimal and a suboptimal behaviour
in the context of risk sensitivity.
Contrary to this prediction,
however,
risk proneness and risk aversion should be feasible 
as well as risk indifference
in the sense that selection against all of them is weak
when $c \approx 0$.

Another relevant  metric is $U/U^*$
where $U$ is the expected fitness of an observed (suboptimal) behaviour
and $U^*$ is the expected fitness for the predicted (optimal) behaviour
\citep{mangel1991adaptive};
in the context of risk sensitivity,
we have
$U=U(\mu,0)$ and $U^*= U(\mu,\sigma)$ on negative budgets
and vice versa on positive budgets.
A caveat of this metric is that it is not affine-invariant.
For instance, the expected fitness offset by $b \ne 0$ fails the invariance, 
i.e.\,$U/U^* \ne \left(U + b\right)/\left(U^* + b\right)$.
Another caveat is that it cannot capture the gradual nature of risk sensitivity on negative budgets
since we have $U/U^* = U(\mu,0) / U(\mu,\sigma) = 0$ 
regardless of $U^*=U(\mu,\sigma) \ne 0$.
This contrasts the case on positive budgets
where $U/U^* = U(\mu,\sigma) / U(\mu,0) =U(\mu,\sigma)$
gradually increases to 1 as mean $\mu$ increases for a given variance $\sigma^2$.

\subsection*{Scalar Utility Theory}
\begin{figure*}[t!]
\centering
\includegraphics[height=8cm]{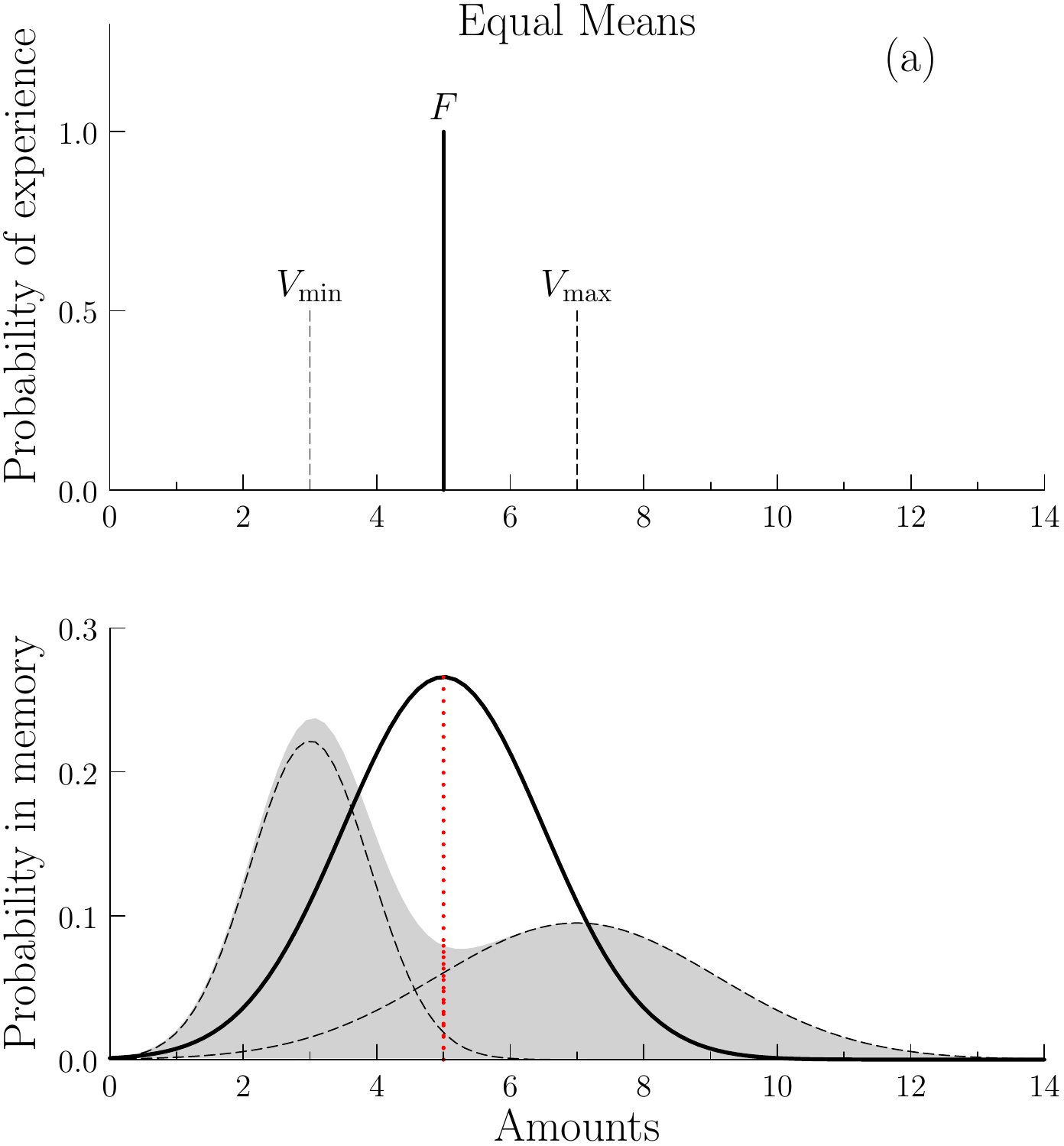} 
\hspace{0.5cm}
\includegraphics[height=8cm]{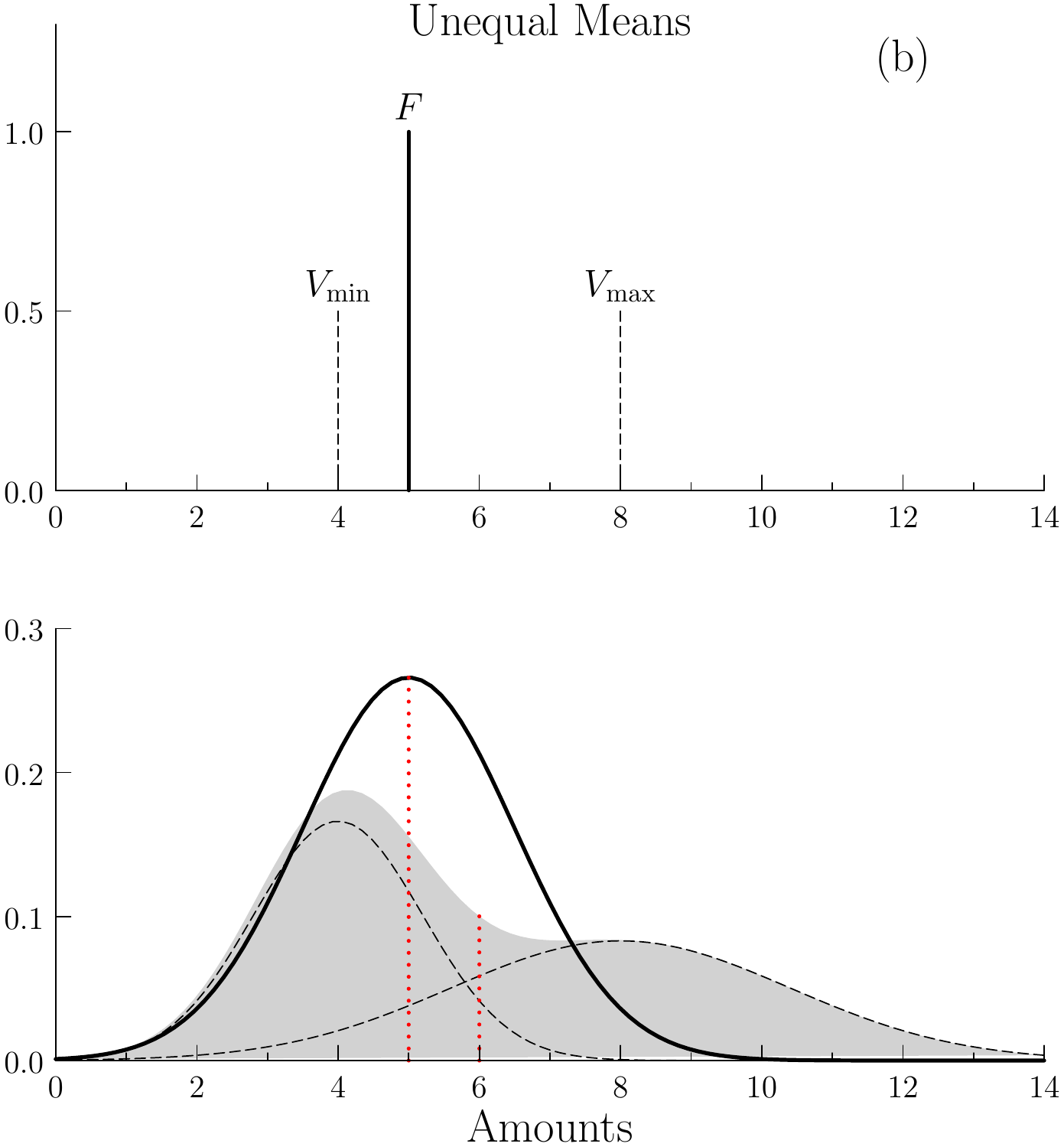} 
\caption{  
SUT (Scalar Utility Theory) for a choice among feeding options of \emph{equal} means vs
\emph{unequal} means.
The memory representation of the constant feeding option $F=5$ 
is shown as a solid curve.
The overall memory representation of the variable option is shown in shading
while each of the small and large amounts is indicated by a dashed curve.
All of the three (normal) distributions have the same coefficient of variation (SD/mean)$=0.3$
which is empirically observed from starlings
\citep{kacelnik1998risky}.
The mean of each representation  is indicated by a dotted line.
	(a)
A variable option $V$ yields
 $V_{\text{min}}=3$ or  $V_{\text{max}}=7$ with equal probability.
The memory representations of the constant option and the variable option have the same mean $=5$.
The mode of the constant option is  5 
and the mode of the variable one is 3;
note that the mode of a 
probability distribution is the value 
at which the distribution has its global maximum.
According to SUT, 
the animal will prefer the constant option since its mode is larger than that of the variable option,
perceiving the amount of the first to be greater than that of the latter and so yielding risk aversion
\citep{kacelnik1998risky}.
SUT typically presents the case of the same mean only,
where both options yield the same expected fitness under the assumption of a linear fitness relation
\citep{kacelnik2013triumphs}.	
	(b)
A variable option $V$ yields 
 $V_{\text{min}}=4$ or $V_{\text{max}}=8$ with equal probability.
The mode of the variable option is  4 
while the mean is 6.
According to SUT, 
the animal will still prefer the constant option 
since the mode of the constant one is larger than that of the variable one
(i.e.\,$5>4$).
However, 
the expected fitness of the constant option is lower than that of the variable option
since  the mean of the constant one is less than that of the variable one (i.e.\,$5<6$)
and SUT assumes a linear fitness relation,
the expected fitness depending only on the mean of food amounts
but no other statistic.
The hypothetical decision process of SUT
yields an erroneous preference of a lower expected fitness,
against which there should be natural selection.
}
\label{fig_SUT}
\end{figure*}

Not viewing risk sensitivity
as having adaptive value per se,~\citet{kacelnik2013triumphs}
instead assert that
it is caused as a side-effect
of perception
and proposes SUT (Scalar Utility Theory)
as an alternative to the conventional budget rule~\citep{kacelnik1998risky}.
SUT consists of two parts,
modelling the animal's knowledge of the food amounts 
and postulating a decision process.
Among feeding options with the same mean but different variances, 
the memory representation of the option with smaller variance has larger skew-sensitive measures
(e.g.\,mode and median)
as a consequence of Weber's law.
If the animal's decision process uses the mode or other skew-sensitive measures
in comparing (the memory representations of) the feeding options,
it is argued that the animal will prefer
the constant option 
since it  provides an `apparently' larger amount of food than the variable option does,
yielding risk aversion;
see Fig.\,\ref{fig_SUT}a.
Predicting only risk aversion
regardless of 
energy budgets,
however,
this side-effect assertion of SUT suffers from numerous caveats from both empirical and theoretical perspectives.

\subsubsection*{Weak empirical support}
As the empirical evidence supporting SUT,
\cite{kacelnik2013triumphs}
refer to a set of 
22 
studies reviewed in~\citet{kacelnik1996risky},
the majority of which report (consistent) risk aversion.
However,
those studies do not manipulate the energy budgets of the subjects.
Since the energy budgets are not manipulated,
the conventional budget rule cannot be fully tested nor fairly compared using these studies.
Risk aversion reported from the studies without budget manipulation does not exclusively support SUT 
in the sense that risk aversion is also predicted by the conventional budget rule
when the animals are on positive budgets.
Even from a perspective of the conventional budget rule,
it is already understood that
risk aversion occurs more often than risk proneness
because the positive budgets are commonly realised levels of energy budgets
\citep{caraco1980foraging}.
The 35 studies 
reviewed in~\citet{kacelnik2013triumphs}
are better suited to fairly test both SUT and the conventional budget rule
using the same data set
since these studies manipulated energy budgets;
additional benefits 
include a larger number of studies 
as well as
an inclusion of
more recent studies.
Among the 35 studies,
8 report consistent risk aversion (as predicted by SUT),
11 of `some amount' 
of risk aversion,
and 
16 of `no' risk aversion.
Only 8/35 support SUT,
which is even lower than 10/35 of supporting the conventional budget rule.
Empirical support of SUT is weaker than that of the conventional budget rule.

\subsubsection*{Description but not explanation}

Animals are assumed to prefer larger amounts of food
\citep{kacelnik2013triumphs}.
Depending on the statistics of the memory representation used for assessing food amounts,
however,
the hypothetical decision process would yield a different choice among feeding options with the same mean and different variances.
If the mean of the memory representation is assumed to be used, it yields no bias to variability of food amounts (i.e.\,risk indifference)
since the animal would `perceive'  the food amounts of the options to be the same.
If the maximum is used, it yields bias in favour of variability (i.e.\,risk proneness)
since the animal would perceive  the amount of the variable option to be  greater;
its positively skewed memory representation
leads to the maximum greater than that of the symmetric distribution of the constant option.
If the mode or median 
is used, it yields bias against variability (i.e.\,risk aversion)
since  the animal would perceive  the  amounts of the constant option to be greater.
What  is the rationale or justification of SUT's assumption
that the animal's decision process would use the mode
(or any other skew-sensitive statistic)
of the memory representations
rather than an alternative statistic such as
the mean or maximum?
If the mode is used in choice process,
it would fit to the empirical data of animals' bias against variability of food amounts
whereas either the mean or maximum would not
\citep{kacelnik1998risky}.
This is hardly an `explanation' of the empirically observed bias,
but  
seems to be no better than
 a `description' of it, 
merely fitting to the data.
If the empirical data showed no bias related to variability, 
according to the logic of SUT,
one would simply 
assume the decision process used the mean
since it would better fit to the data
than other statistics would.
If the data showed a bias in favour of variability,
one would just assume to use the maximum
or any other statistic that would yield the bias,
according to the fitting-to-data logic of SUT.

\subsubsection*{Fitness consequence of erroneous preferences}
Another caveat of the usage of the mode
or any other skew-sensitive statistic
is 
the functional problem
of preferring an option of lower fitness. 
Unlike other relevant theories in biology, psychology and economics,
SUT assumes fitness (or utility) as a `linear' function of 
food amounts
\citep{kacelnik1998risky}.
It is well known that, under  the linear relations, 
only the mean of food amounts
 has an effect on the mean of fitness (i.e.\,expected fitness),
but variance or any other variability-sensitive statistic
does not
\citep{houston1999models,
stephens1987foraging}.
In any theory that assumes a linear fitness function as SUT does,
the animal's decision process 
is thus expected to use and compare the mean 
 of 
food amounts
(i.e. preferring a feeding option of higher mean).
However,
SUT assumes that the decision process uses a skew-sensitive statistic
such as the mode
(i.e. preferring an option of higher mode).
This is not a merely logical issue,
but has a serious fitness consequence
as well.
If the decision process is based on a skew-sensitive statistic
(e.g.\,mode)
as assumed by SUT,
it can yield an erroneous preference of an option
whose expected fitness is lower than that of the alternative.
Even though the mean amount of a variable feeding option is higher than that of a constant one
(and thus the expected fitness of the first is higher than that of the latter),
for example,  
the animal that 
makes a choice based on the mode
will prefer the constant one 
(of the lower expected fitness)   
if the mode of the variable one is lower than that of the constant one due to the right-skew of the memory representation of the variable one;
see Fig.\,\ref{fig_SUT}b. 
Note that 
this problematic preference of lower fitness
is admitted for
an earlier version of SUT
based on a single sampling of memory representation
\mbox{\citep{kacelnik1998risky}}.
However,
the problem persists
with the later versions of SUT
based on a skew-sensitive statistic;
any decision process
that is based on a statistic other than the mean
yields the problematic preference
although the conditions where the problem occurs may differ,
depending on the statistic.
There should be 
natural 
selection against the erroneous decision process that uses 
a
skew-sensitive statistic.
For SUT to be a genuinely mechanistic model
predicting risk aversion,
it should justify
the usage of the mode (or any skew-sensitive statistic
such as the median or even the single sampling)
on mechanistic grounds
rather than merely assuming it.
For instance, 
suppose that (an algorithm or neural mechanism of) computation of the mode of a distribution is significantly simpler than that of the mean such that it would even compensate the negative consequence due to the mode-based choice,
then it would make a good case for using the mode.
To our knowledge,
unfortunately,
SUT seems to lack of any related attempts
\citep{kacelnik2013triumphs, kacelnik1998risky}.

In summary,
even though Weber's law
as a universal principle of psychophysics
is empirically well supported,
it merely supports one component of SUT,
i.e.\,a Weberian memory representation of food amounts
\citep{bateson1995accuracy}.
However,
it has little to do with supporting the  key component of SUT,
i.e.\,the hypothetical decision process based on skew-sensitive statistics
to yield only risk aversion,
which suffers from numerous caveats.

\subsection*{Comparison to Hybrid Explanations} 
Contrary to SUT that predicts only risk aversion,
one can accommodate both risk aversion and risk proneness 
to be compatible with the conventional budget rule
even though Weber's law is applied
\citep{shafir2000risk}.
While revealing a gradual nature of risk sensitivity from a meta-analysis based on regressions,
\citet{shafir2000risk} concludes that
the direction of preference is determined by functional considerations (complying with the conventional budget rule)
and the strength of preference by perceptual considerations (complying with Weber's law).
When risk aversion is predicted by the conventional budget rule,
for instance,
the regression coefficient is positive, indicating that an increase in relative variability yields stronger risk aversion.
According  to Weber's law,
relative variability or the coefficient of variation CV$=\frac{\sigma}{\mu}$ correlates with capability of perceptual discrimination 
between a variable and a constant feeding 
option.
Thus, the stronger risk aversion
or the increasing preference of a constant option on positive budgets
with increasing CV of the variable feeding option
is explained by perceptual considerations
in the sense that
perceptual discrimination of a constant option (i.e.\,the optimal choice)
from a variable one (i.e.\,the suboptimal choice)
improves
as the difference in CV between the options increases 
\citep{shafir2000risk}. 
On the other hand,
it is argued that
the functional considerations based on
the conventional budget rule  
are unable to
predict this changing level of risk sensitivity between experimental conditions
\citep{shafir2000risk, shafir2005caste}.
According to the gradual budget rule,
however,
even the strength of preference can be explained by functional considerations as well.
For a given standard deviation $\sigma \ne 0$,
as mean $\mu$ decreases on positive budgets, 
the difference in 
expected fitness
between the feeding options  increases
(i.e.\,selection pressure for risk aversion strengthens)  
while the difference in CV increases as well 
since CV$=\frac{\sigma}{\mu}$ negatively correlates with mean $\mu$.
A positive correlation between strength of risk aversion and CV 
thus stems from functional considerations (complying with the gradual budget rule) as well.

\subsection*{Empirical Tests of the Gradual Budget Rule}

While the aforementioned meta-analysis supports the gradual budget rule,
it is based on different groups of subjects under different energy budgets.
To empirically test the gradual budget rule within the same individuals,
one needs to involve a wide range of parameters.
Instead of a single 
or a limited range of reserve mean
on positive budgets, 
for instance,
a wider range  should be tested out
since the gradual budget rule not only predicts a richer set of risk sensitivity
than the conventional budget rule does,
but also complies with the latter,
depending on the reserve mean.
For example,
hummingbirds not only become more risk-averse on positive budgets
as the reserve mean gets closer to the survival threshold,
but also show more risk proneness on positive budgets
when the reserve mean gets further away from the threshold
\citep{hurly2003twin}.
European starlings  become more risk-averse on positive budgets
as ambient temperature decreases, 
the latter of which is equivalent to raising the survival threshold, 
yielding a reserve mean closer to it
\citep{bateson2002context}.

\section*{Conclusions}
We proposed the gradual budget rule, which normatively reflects the gradual nature of risk sensitivity
and encompasses the energy budget rule as a special case.
The gradual budget rule predicts significant selection pressure 
 for the risk sensitivity complying with the energy budget rule in a limited range of parameters
 and less significant pressure in the remaining range,
the latter of which can allow for behaviours not complying with the energy budget rule
and
explains some of the empirical findings that used to be counted against the energy budget rule
as well as the adaptive value of risk sensitivity.

\section*{Acknowledgements}
This work was supported in part by a grant funded by the Wales European Funding Office of the Welsh Government for the Wales Centre for Behaviour Change, Bangor University. 


\bibliographystyle{apalike}


\newpage

\end{document}